\documentclass[reprint,prd,letterpaper,superscriptaddress,amsmath,amssymb,aps,showpacs,twocolumn,showkeys] {revtex4}

\usepackage{graphicx}
\usepackage{dcolumn}
\usepackage{todonotes}
\usepackage{amsmath}
\usepackage{bm}
\usepackage{siunitx}
\usepackage{subfigure}
\usepackage{pifont}
\usepackage{xcolor}

\usepackage[implicit=true, debug=false,
plainpages=false, colorlinks=true,
linkcolor=black, urlcolor=black, citecolor=black, filecolor=black,
hyperindex=true]{hyperref}

\newcommand{\be}{\begin{equation}}
\newcommand{\ee}{\end{equation}}
\newcommand{\bq}{\begin{eqnarray}}
\newcommand{\eq}{\end{eqnarray}}

\def\({\left(}
\def\){\right)}

\begin{document}
\bibliographystyle{apsrev4-1}
\title{Exploring the Latest Pantheon SNIa Dataset by Using Three Kinds of Statistics Techniques}

\author{Shuang Wang\footnote{Corresponding author}}

\email{wangshuang@mail.sysu.edu.cn}
\affiliation{School of Physics and Astronomy, Sun Yat-Sen University, Guangzhou 510297, P. R. China}

\author{Xiaolin Luo}
\affiliation{School of Physics and Astronomy, Sun Yat-Sen University, Guangzhou 510297, P. R. China}

\date{\today}

\begin{abstract}
  In this work, we explore the cosmological consequences of the latest Type Ia supernova (SN Ia) data-set, Pantheon, by adopting the $wCDM$ model.
  The Pantheon data-set is the largest SN Ia samples till now, which contains 1048 supernovae on the redshift range $0 < z < 2.3$.
  Here we take into account three kinds of SN Ia statistics techniques, including:
  1. magnitude statistics (MS), which is the traditional SN Ia statistics technique;
  2. flux statistics (FS), which bases on the flux-averaging (FA) method;
  3. improved flux statistics (IFS), which combines the advantages of MS and FS.
  It should be mentioned that, The IFS technique need to scan the $(z_{cut},\Delta z)$ parameters plane,
  where $z_{cut}$  and $\Delta z$ are redshift cut-off and redshift interval of FA, respectively.
  The results are shown as follows.
  (1) Using SN data-set only, the best FA recipe for IFS is $(z_{cut},\Delta z)=(0.1,0.08)$;
  (2) Comparing to the old SN data-set, JLA, adopting Pantheon data-set can reduce the $2\sigma$ error bars of equation of state $w$ by 38\%, 47\% and 53\% for MS, FS and IFS, respectively;
  (3) FS gives closer results to other observations, such as Baryon acoustic oscillations and Cosmic microwave background;
  (4) Compared with FS and IFS, MS more favors a Universe that will end in a ``big rip''.


\end{abstract}

\maketitle

\

\section{Introduction}\label{intro}
Type Ia supernova (SN Ia), which is a sub-category of cataclysmic variable stars \citep{Hillebrandt:2000ga},
had played a crucial role in discovering the cosmic acceleration \citep{Riess:1998cb,Perlmutter:1998np}.
So far, SN Ia is still one of most useful and powerful tools to investigate the nature of dark energe (DE) \citep{Peebles:2002gy,Padmanabhan:2002ji,Copeland:2006wr,Frieman:2008sn,Caldwell:2009ix,
Li:2011sd,Bamba:2012cp,Li:2012dt,Wang:2016och}.

In recent ten years, many high quality supernova data-set have been released, such as ``Union'' \citep{Kowalski:2008ez},
``Constitution'' \citep{Hicken:2009dk}, ``SDSS'' \citep{Kessler:2009ys}, ``Union2'' \citep{Amanullah:2010vv}, ``SNLS3'' \citep{Conley:2011ku},
``Union2.1'' \citep{Suzuki:2011hu} and ``JLA'' \citep{Betoule:2014frx}.
In 2018, the latest ``Pantheon'' data-set \citep{Scolnic:2017caz},
which contains 1048 SNIa at the redshift range $0 < z < 2.3$, was released.
Using the Pantheon sample, Scolnic et al. had given the observational constraints on the $w CDM$ and $CPL$ models \citep{Scolnic:2017caz}.

On the other hand , along with the rapid growth of the number of SN Ia discovered,
the studies on the systematic uncertainties of SN Ia have drawn more and more attention.
It has been proved that \citep{Weinberg:2012es}, the classic SN Ia statistics method
(hereafter we will call it ``magnitude statistic'' (MS))
suffers from various systematic uncertainties ,
such as the calibration errors \citep{Conley:2011ku}, the host-galaxy extinction \citep{Kelly:2009iy,Lampeitl:2010zx,Sullivan:2010mg}, the gravitational lensing \citep{Frieman:1996xk,Wang:1999me}, different light-curve fitters \citep{Hu:2015opa} and the redshift evolution of SN  color luminosity parameter $\beta$ \citep{Wang:2013yja,Mohlabeng:2013gda,Wang:2013tic,Wang:2014oga,Wang:2014fqa,2015SCPMA..58a..17W,Li:2016dqg}.
Therefore, the control of the systematic uncertainties of SNIa have become one of the biggest challenges in SN cosmology.

In order to reduce the systematic uncertainties of SNIa, some interesting statistics techniques of SN Ia are proposed in the literature.
For examples, in 2000, Wang proposed a new analysis technique, called flux-averaging (FA),
to reduce the systematic errors caused by the weak lensing effect of SNIa \citep{Wang:1999bz}.
The FA technique focus on the observed flux data of SN Ia,
and then average these flux data at some redshift bins with same width.
Hereafter, we will call this statistics method of SN Ia as ``flux statistic'' (FS).
The FS can reduce several systematic uncertainties of SN Ia \citep{Wang:2005yaa,Wang:2009sn,Wang:2011sb},
but it will lead to larger error bars of model parameters.
In 2013, One of the present authors and Wang \citep{Wang:2013yja} proposed an improved version of flux-averaging.
This new statistics method combines the advantages of MS and FS,
and thus can reduce the systematic uncertainties and the error bars of model parameters at the same time \citep{Wang:2016bba,Wen:2017aaa,Luo:2018yvq}.
Hereafter, we will call this latest statistics method of SN Ia as ``improved flux statistics''  (IFS).

In this work, we will explore the Latest Pantheon SN Ia data-set by using all the three statistic techniques of SN Ia (i.e. MS, FS and IFS).
It should be emphasized that, in the previous studies about the Pantheon samples \citep{Riess:2017lxs,Andrade:2018eta,Deng:2018jrp,Abbott:2018wog}, only MS was taken into account.
On the other hand, in our previous studies \citep{Wang:2016bba,Wen:2017aaa,Luo:2018yvq},
The cosmology-fits are always performed by combining SN Ia samples with other observations, such as Cosmic microwave background(CMB) and Baryon acoustic oscillations (BAO).
In this work, we will mainly focus on the cosmological constraints given by the Pantheon SN Ia Data-set alone.

The paper is organized as follows.
In Section.\ref{data}, we will introduce the methodology used in this work.
In particular, we will show how to calculate the $\chi^2$ function of SN Ia data, for the case of adopting MS, FS, IFS, respectively.
In Section.\ref{result}, we will show the results of our studies.
We will discuss the differences between cosmological consequences given by Pantheon data-set and by pervious SN samples,
the differences among the cosmological constraints given by MS, FS, IFS,
as well as the ultimate fate of the Universe.
Finally, the discussion and conclusion are shown in Section.\ref{conclu}

\section{Methodology}\label{data}

In this section, we introduce how to calculate the $\chi^2$ function of SN Ia data, for the case of adopting MS, FS, IFS, respectively.

\subsection{Magnitude statistics}

As shown in Ref.\citep{Scolnic:2017caz}, adopting MS,
the $\chi^2$ function of SN Ia data can be expressed as
\be
\chi^2 = \Delta\mu^T\cdot {\bf Cov}^{-1}\cdot\Delta\mu.
\ee

Here the $\Delta\mu\equiv\mu_{obs}-\mu_{th}$,
where $\mu_{obs}$ is the observational distance modulus of SN, given by \citep{Scolnic:2017caz}
\be\label{eqt:muobs}
\mu_{obs}=m_B-M+\alpha X_1-\beta C+\Delta_M+\Delta_B,
\ee
where $m_B$ is the observed peak magnitude in the rest frame of the B band,
$M$ is the absolute B-band magnitude of a fiducial SNIa,
$\alpha$ is the coefficient of the relation between luminosity and stretch,
$X_1$ describes the time stretching of the light curve,
$\beta$ is the coefficient of the relation between luminosity and color,
and $C$ describes the supernova color at maximum brightness.
Furthermore, $\Delta_M$ is a distance correction based on the host-galaxy mass of the SNIa and $\Delta_B$ is a distance correction based on predicted biases from simulation.

The theoretically distance modulus of SN Ia $\mu_{th}$ can be expressed as
\be
\mu_{th} = 5 \log_{10}\bigg[\frac{d_L(z_{hel},z_{cmb})}{Mpc}\bigg] + 25.
\ee
Here $z_{cmb}$ is the CMB restframe redshift, $z_{hel}$ is the beliocentric redshift,
and $d_L$ is the luminosity distance of  SN Ia, given by
\be
d_l(z_{hel},z_{cmb}) = (1+z_{hel})r(z_{cmb}).
\ee
$r(z)$ is given by
\begin{equation}
r(z) = c H_0^{-1}\int_0^z\frac{dz'}{E(z')},
\end{equation}
where $c$ is the speed of light, $H_0$ is the current value of the Hubble parameter $H(z)$, and $E(z)\equiv H(z)/H_0$.

For simplicity, in this work we only consider the $w CDM$ model (i.e. DE equation-of-state(EOF) $w$ is a constant parameter ) in a flat Universe. Based on the Friedmann equation, we can get
\be
E(z) = \sqrt{\Omega_r(1+z)^4+\Omega_m(1+z)^3+\Omega_{de}(1+z)^{3(1+w)}},
\ee
where $\Omega_r,\Omega_m$ and $\Omega_{de}$ represent the current fractional densities of radiation, matter and dark energy, respectively.
And the radiation density parameter $\Omega_r$ is given by Ref.\citep{Wang:2013mha},
\be
\Omega_{r}=\Omega_{m}/(1+z_{\rm eq}),
\ee
where $z_{\rm eq}=2.5\times10^4\Omega_{m}h^2(T_{\rm cmb}/2.7\,{\rm K})^{-4}$, $T_{\rm cmb}=2.7255\,{\rm K}$, and $h$ is the reduced Hubble constant.
In the case of only adopting SN samples, we set the radiation density parameter $\Omega_r=0$.

In addition, ${\bf Cov}$ is the total covariance matrix, which is given by
\be
{\bf Cov} = D_{stat} + C_{sys},
\ee
where the statistical matrix $D_{stat}$ only has the diagonal components, it includes the distance error of each SNIa as follow
\be
\sigma^2 = \sigma^2_N + \sigma^2_{Mass} + \sigma^2_{\mu-z} + \sigma^2_{lens} + \sigma^2_{int} + \sigma^2_{Bias},
\ee
where $\sigma^2_N$ is the photometric error of the SNIa distance, $\sigma^2_{Mass}$ is the distance uncertainty from the mass step correction, $\sigma^2_{\mu-z}$ is the uncertainty from the peculiar velocity uncertainty and redshift measurement uncertainty in quadrature, $\sigma^2_{lens}$ is the uncertainty fron stochastic gravitational lensing, $\sigma^2_{int}$ is the intrinsic scatter, and $\sigma^2_{Bias}$ is the uncertainty from the distance bias correction.
Furthermore, $C_{sys}$ is the systematic covariance for each SNIa.
One can find the more details about the uncertainty matrix ${\bf Cov}$ in Ref.\cite{Conley:2011ku}.

\subsection{Flux statistics}
FA divides the whole redshift region into some bins with the same width.
The segment points of various bins are $z_i=\Delta z\cdot i$, where $\Delta z$ is the width of each bin and $i=1,2,3,\ldots,n$.

As shown in Ref.\citep{Wang:1999bz}, adopting FS,
the $\chi^2$ function of SN Ia data can be expressed as
\be
\chi^2=\sum_{ij}\Delta\overline{\mu}(\overline{z}_i)Cov^{-1}
[\overline{\mu}(\overline{z}_i),\overline{\mu}(\overline{z}_j)]
\Delta\overline{\mu}(\overline{z}_j),
\ee
where
\be
\Delta\overline{\mu}(\overline{z}_i)\equiv\overline{\mu}^{obs}(\overline{z}_i)
-\overline{\mu}^p(\overline{z}_i|{\bf s})
\ee
with the average redshift $\overline{z}_i = \frac{1}{N_i}\sum_{l=1}^{N_i}z_{l,cmb}^{(i)}$ in the $i$-th bin.

The observational flux-averaged distance modulus is calculated by
\be
\overline{\mu}^{obs}(\overline{z}_i) = -2.5 \log_{10}\overline{F}(\overline{z}_i)+25,
\ee
where $\overline{F}(\overline{z}_i) = \overline{\cal L}^i/d_L^2(\overline{z}_i|{\bf s})$ is
the binned flux in the $i$-th redshift bin.
Here, $\{\bf s\}$ represents a set of cosmological parameters.
The the average value of $\{{\cal L}(z_{cmb})\}$ in each redshift bin $i$ is given by
\be
\overline{\cal L}^i = \frac{1}{N_i}\sum_{l=1}^{N_i}{\cal L}_l^i(z_{l,cmb}^{(i)}),
\ee
where the ``absolute luminosities'' $\{{\cal L}(z_{cmb})\}$ is
\be
{\cal L}(z_{cmb})\equiv d_L^2(z_{cmb}|{\bf s})F(z_{cmb})
\ee
with the ``fluxs'' distance modulus
\be
F(z_{cmb})\equiv10^{-(\mu_0^{obs}(z_{cmb})-25)/2.5} = \bigg(\frac{d_L^{obs}(z_{cmb})}{Mpc}\bigg)^{-2}.
\ee

On the other hand, the theoretical prediction is given by
\be
\overline{\mu}^p(\overline{z}_i) = -2.5\log_{10}F^p(\overline{z}_i)+25
\ee
with $F^p(\overline{z}_i|{\bf s})=(d_L(\overline{z}_i|{\bf s})/Mpc)^{-2}$.

The covariance matrix of $\overline{\mu}(\overline{z}_i)$ and $\overline{\mu}(\overline{z}_j)$ is calculated as
\be
\begin{aligned}
Cov[\overline{\mu}(\overline{z}_i),\overline{\mu}(\overline{z}_j)] &=
\frac{1}{N_i N_j \overline{L}^i \overline{L}^j}\\ &\sum_{l=1}^{N_i}\sum_{m=1}^{N_j}
{\cal L}(z_l^{(i)}){\cal L}(z_m^{(j)})
\langle\Delta\mu_0^{obs}(z_l^{(i)})\Delta\mu_0^{obs}(z_m^{(j)})\rangle,
\end{aligned}
\ee
where $\langle\Delta\mu_0^{obs}(z_l^{(i)})\Delta\mu_0^{obs}(z_m^{(j)})\rangle$ is the covariance of the measured distance moduli if the $l$-th SNIa in the $i$-th redshift bin, and the $m$-th SNIa in te $j$-th redshift bin.

For more details about the FA technique, see the Ref.\citep{Wang:1999bz}.
\subsection{Improved flux statistics}
IFS introduces a new parameter, i.e. redshift cut-off $z_{cut}$.
For the case of $z<z_{cut}$, the $\chi^2$ function is calculated by using MS;
for the case of $z\geq z_{cut}$, the $\chi^2$ function is calculated by using FS.
It means that,
\be
\chi^2_{IFS} = \chi^2_{MS}(z<z_{cut})+\chi^2_{FS}(z\geq z_{cut}).
\ee

Comparing to MS, IFS introduces two new parameters, i.e. redshift cut-off $z_{cut}$ and the width of redshift bin $\Delta z$.
Here we require that $z_{cut}=0.1 \cdot i$, $i=0,1,2,\ldots,8$; while $\Delta z=0.01 \cdot j$, $j=4,5,6,\ldots,11$.

Based on the JLA samples, Ref.\citep{Wang:2016bba} scanned the whole $(z_{cut},\Delta z)$ plane,
and found that $(z_{cut},\Delta z)=(0.6,0.06)$ will give the tightest DE constrains.
But this result was obtained by using a combined observational data, which includes CMB, BAO and SNIa.
In this work, we mainly focus on the SN Ia data.
Therefore, using Pantheon sample alone, we will scan the parameter space of $(z_{cut},\Delta z)$.

\subsection{Other observational data}

In addition to the SN Ia samples, some other cosmological observations,
such as CMB \citep{Bennett:2003ba,Komatsu:2008hk,Wang:2013mha} and BAO \citep{Tegmark:2003ud,Tegmark:2006az,Hu:2015ksa},
also play important roles in exploring the nature of DE.
Therefore, for comparison, we also take CMB and BAO data into account.

In this work, for CMB, we use the distance priors data extracted from Planck 2015 \citep{Ade:2015rim}.
For BAO, we adopt the data from BOSS DR12 \citep{Alam:2016hwk}, which provides 6 data points of $H(z)$ and $D_A(z)$ at $z=0.38,0.51$ and $0.61$.
For more details about the way of calculating $\chi^2$ of CMB and BAO, see Ref.\cite{Wen:2017aaa}.
\section{Cosmology-fit results}\label{result}
The cosmology-fits of this work are performed by using COSMOMC package \citep{Lewis:2002ah}.
Moreover, to access the ability of constraining DE for various SN Ia statistics techniques,
we also take into account the quantity figure of marit (FoM) \citep{Albrecht:2006um,Wang:2008zh},
which is the inverse of the area enclose by the $2\sigma$ confidence level (CL) contour of $(w,\Omega_m)$, for the $wCDM$ model.
Therefore,
\be\label{eqt:fom}
FoM=\frac{1}{\sqrt{det\, Cov(f_1,f_2,f_3,\cdots)}},
\ee
where $Cov(f_1,f_2,f_3,\cdots)$ is the covariance matrix of the chosen set of DE parameters.

In this section, first of all, we discuss the best recipe for IFS. Then, we compare the cosmological consequence of Pantheon data-set with the results of JLA data-set.
Finally, we compare the results of MS, FS and IFS, respectively.
\subsection{Searching the best FA recipe for IFS}
In this subsection, we scan the $(z_{cut},\Delta z)$ plane to find the best FA recipe for IFS.
As mentioned above, we require that $z_{cut}=0.1 \cdot i$, $i=0,1,2,\ldots,8$; while $\Delta z=0.01 \cdot j$, $j=4,5,6,\ldots,11$.
For each set of $(z_{cut},\Delta z)$, we perform a MCMC analyse by using the $w CDM$ model. Then, we compute the corresponding values of FoM, which are given by Eq.\ref{eqt:fom}.

\begin{figure*}[htbp]
      \centering

      \includegraphics[width=12cm,height=8cm,trim=0 50 120 80,clip]{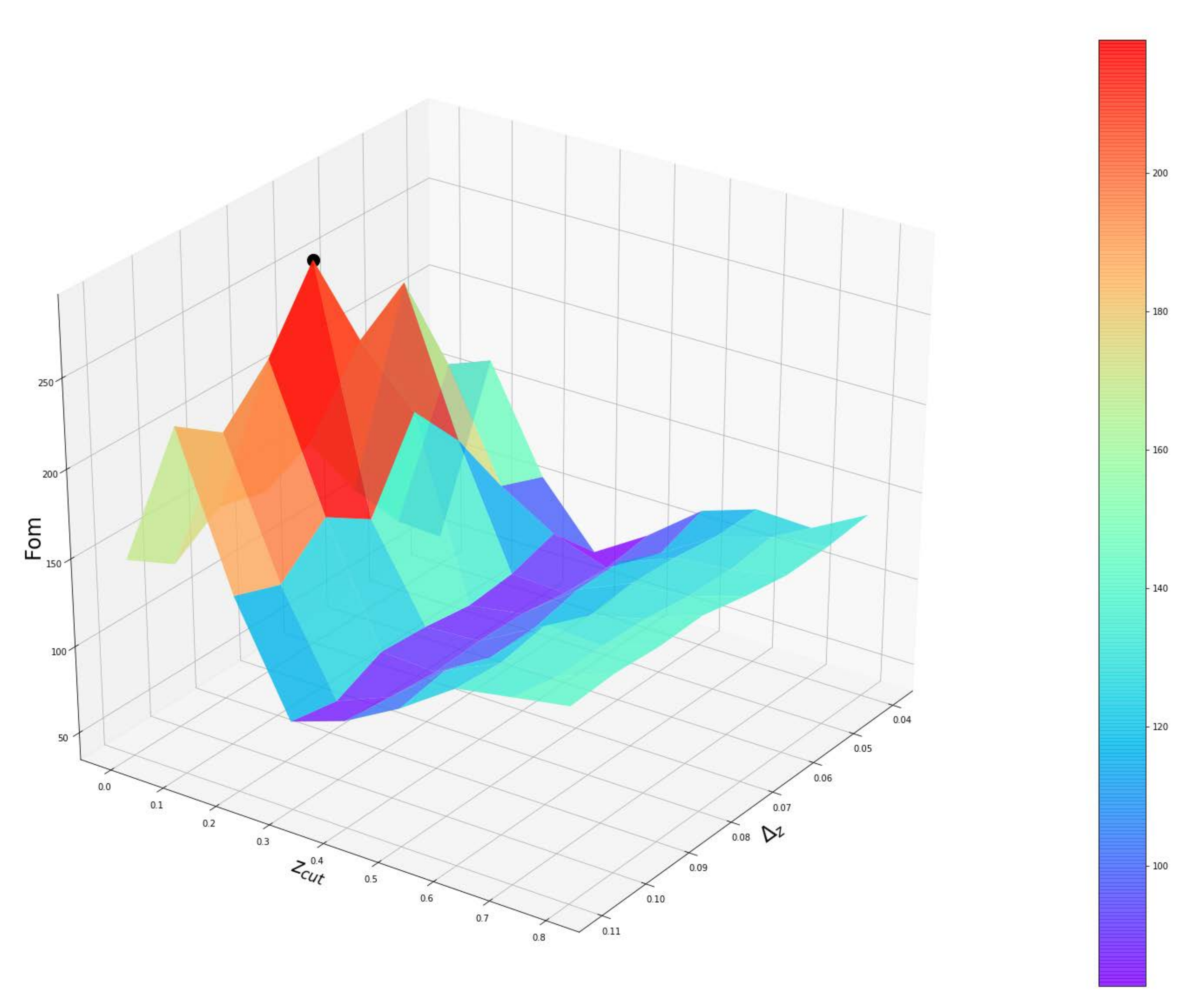}
      \caption{The 3D graph about the values of FoM, which are given by different sets of $(z_{cut},\Delta z)$, for $wCDM$ model. The black dot represents the best FA recipe $(z_{cut},\Delta z)=(0.1,0.08)$ for IFS, which gives the $FoM=278.34$.}
      \label{fig:scan}
      \end{figure*}

The 3D graph about the values of FoM, given by different sets of $(z_{cut},\Delta z)$,
is shown in Fig.\ref{fig:scan}. It's clear that different values of $(z_{cut},\Delta z)$ will give different FoM. Based on this figure, we find that the best FA recipe of IFS is $(z_{cut},\Delta z)=(0.1,0.08)$ with the FoM $=278.24$ (denoted by a black dot). Hereafter, we use this recipe for all the IFS technique.

\begin{figure}[htbp]
      \centering

      \includegraphics[width=7cm,height=5cm]{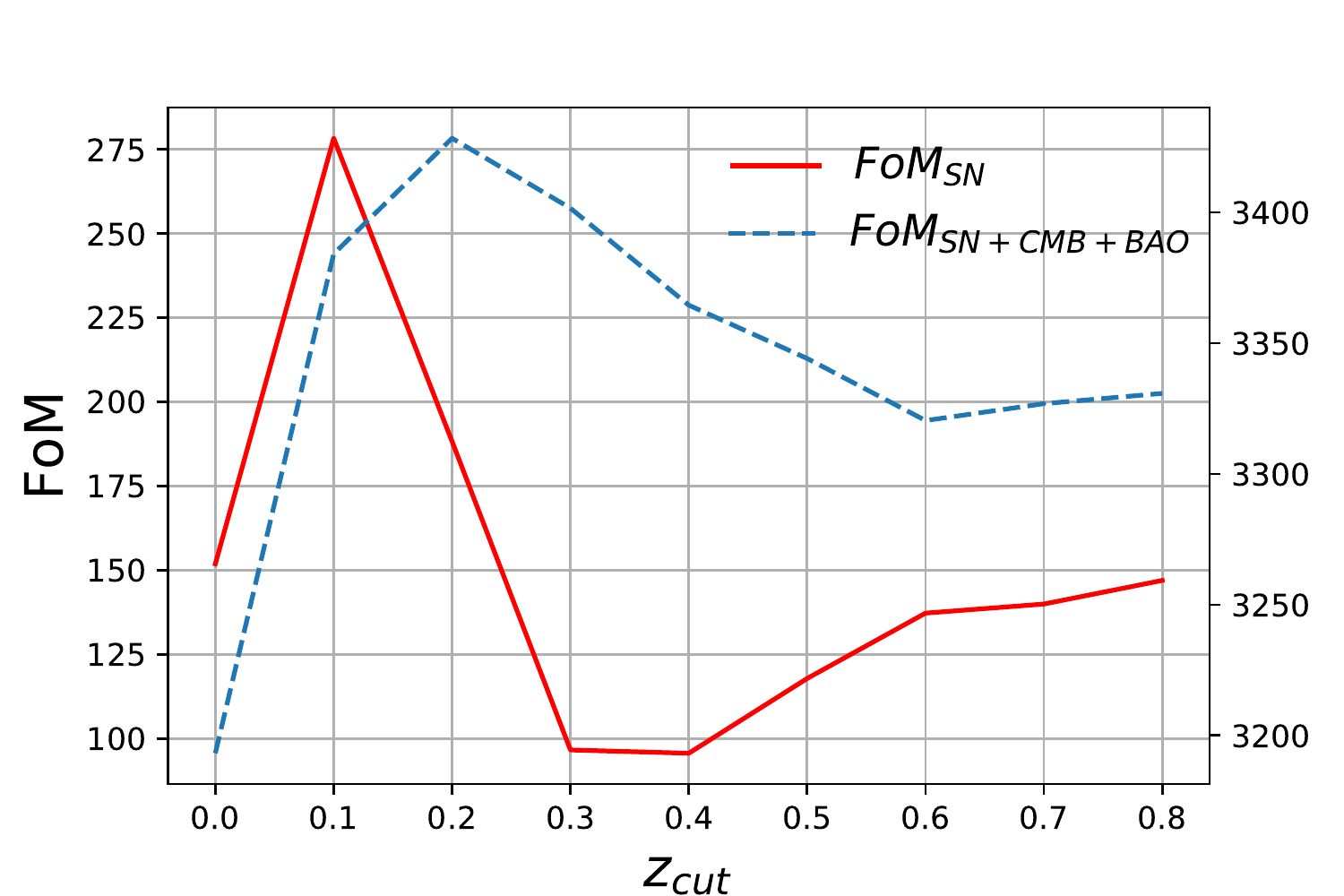}

      \caption{The FoM values given by different $z_{cut}$, for $wCDM$. The solid red line represents the results constrained only by SN samples. The blue dashed line represents results constrained by the combined SN+BAO+CMB data.}
      \label{fig:cut}
      \end{figure}

Form the above figure, one can see that varying $z_{cut}$ will produce larger influence on the value of FoM than changing $\Delta z$.
In Fig.\ref{fig:cut}, we give the results of FoM given by different $z_{cut}$.
The solid red line denotes the results given by using the SN data alone,
while the dashed blue line represents the results given by using the combined SN+CMB+BAO data.
For the case of using the SN data alone, the values of FOM rapidly decrease at the region $z_{cut}>0.1$.
For the case of using the combined SN+CMB+BAO data, $z_{cut}=0.2$ will yield the maximal value of FOM.
In other words, using the SN data alone will give a smaller  $z_{cut}$.

\subsection{The differences between the cosmological consequences of Pantheon and JLA}
In this subsection, we compare the differences between the cosmological consequences of Pantheon and JLA.
For complete analysis, all the three statistics techniques, including MS, FS and IFS, are taken into account.

\begin{figure*}[htbp]
    \centering

      \includegraphics[width=5cm,height=5cm,trim=120 270 120 270,clip]{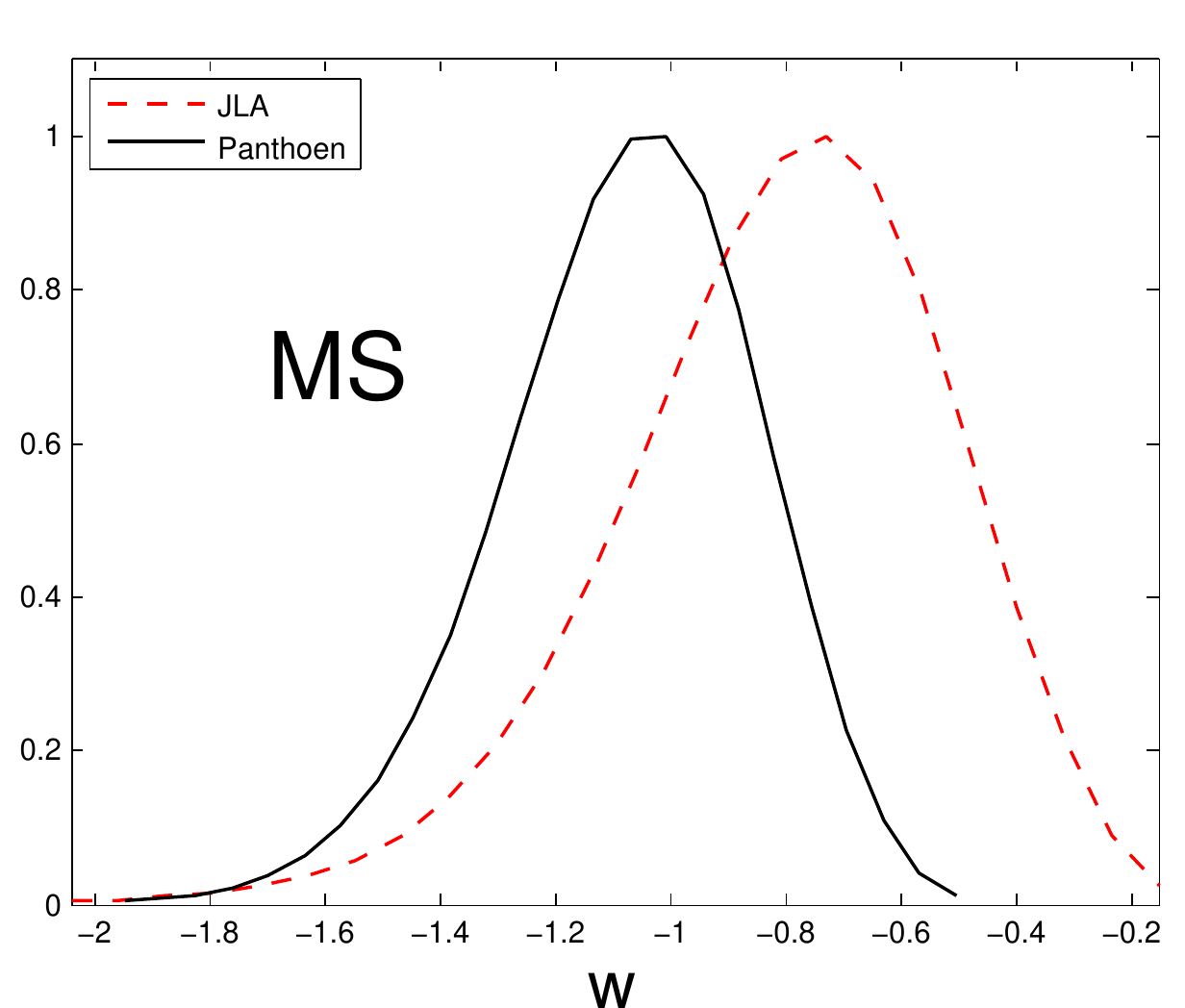}
      \includegraphics[width=5cm,height=5cm,trim=120 270 120 270,clip]{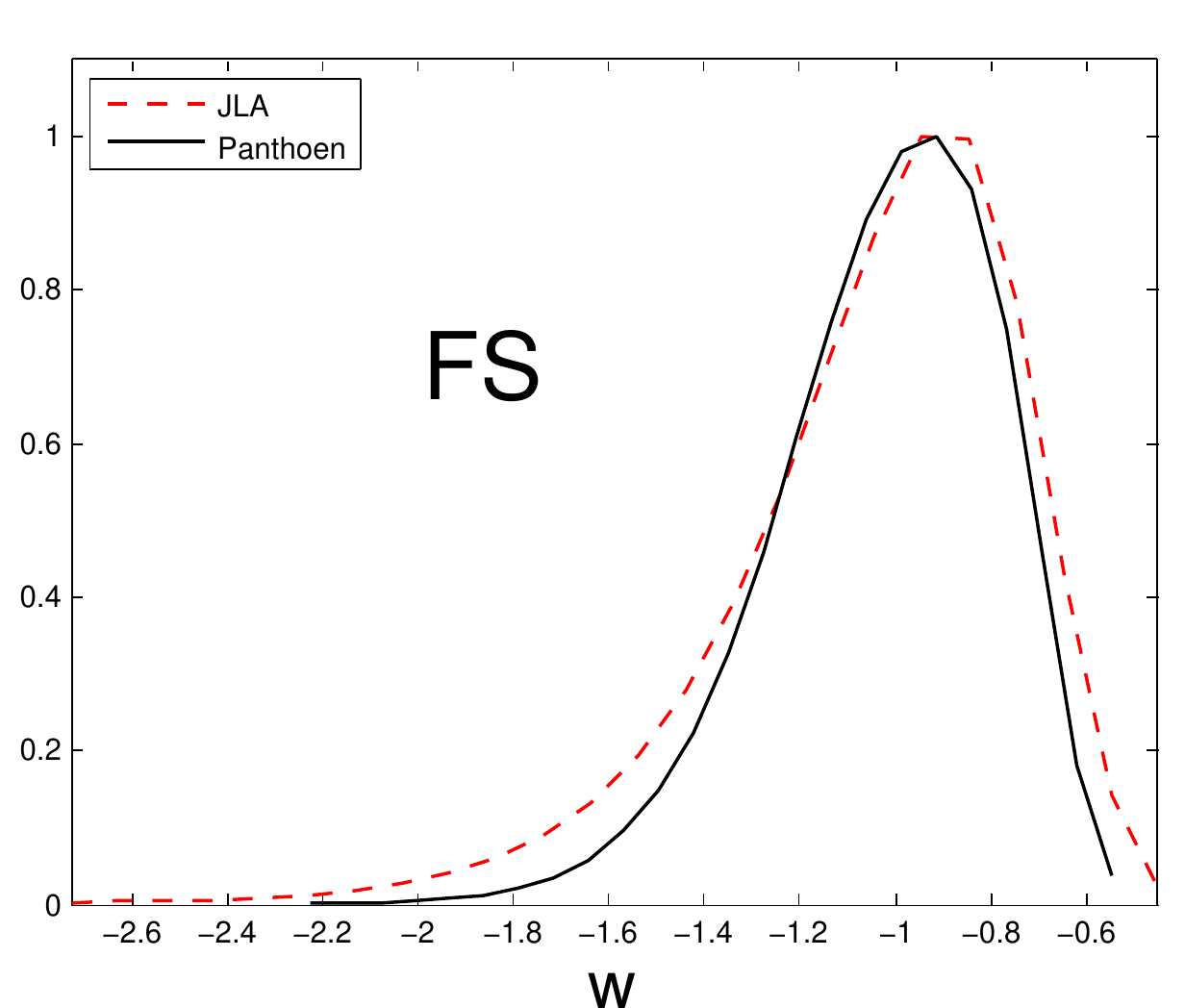}
      \includegraphics[width=5cm,height=5cm,trim=120 270 120 270,clip]{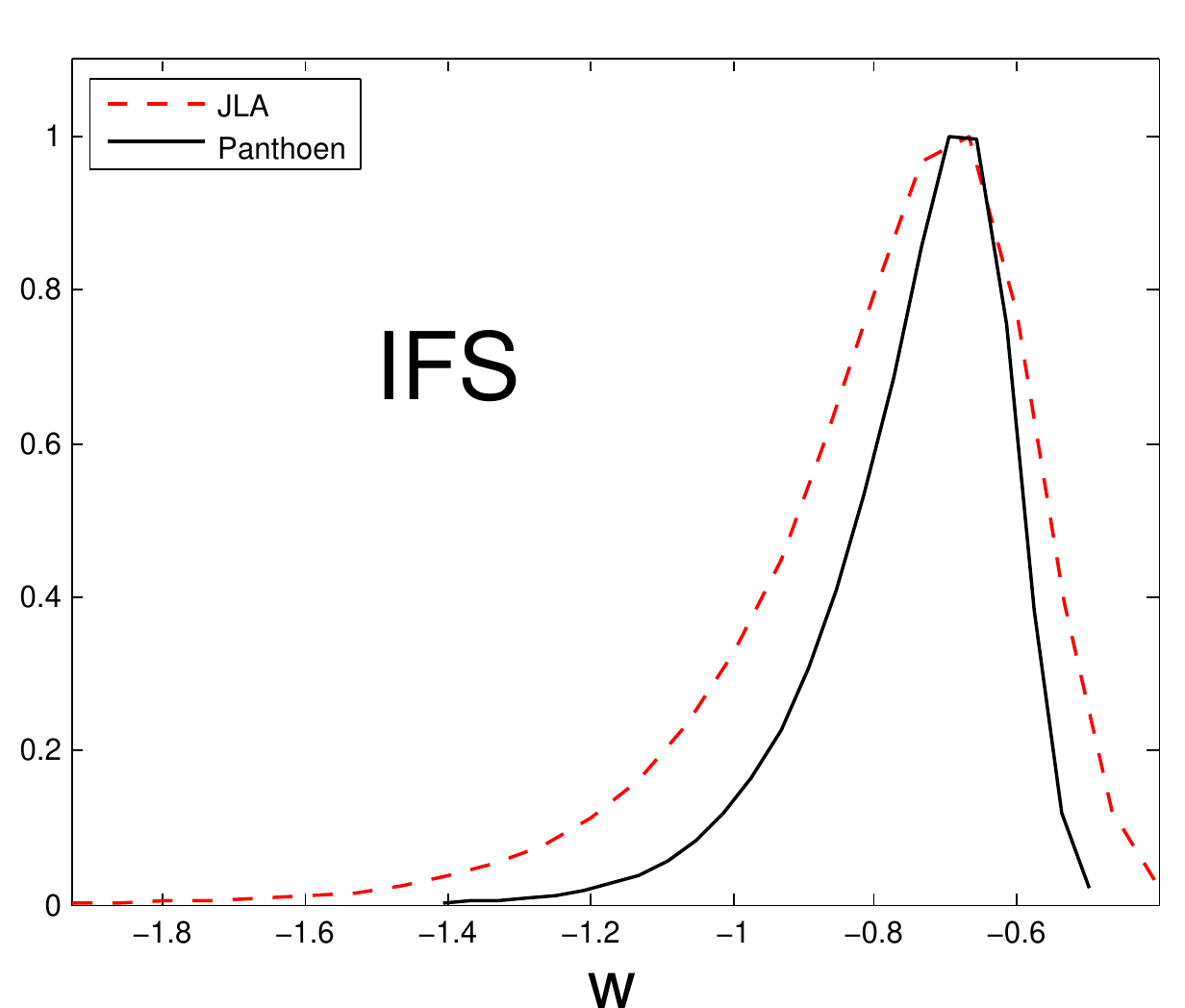}
    \caption{The 1D marginalized probability distributions of $w$ for three SN Ia statistics techniques. The solid black lines denote the result from Pantheon, and the dashed red lines represent the result from JLA. For all the statistics techniques, the new data-set, Pantheon , gives tighter constraint on $w$ than that constrained by JLA data-set.}\label{fig:three}
    \end{figure*}

In Fig.\ref{fig:three}, we present the 1D marginalized probability distributions of $w$ , which is produced by the Pantheon and JLA data-set, respectively.
One can find that, for the case of using MS technique, the results of $w$ given by the Pantheon and the JLA data are quite different.
for the case of using FS and IFS, the Pantheon and the JLA data will give the similar results of $w$.

More details are shown in Table.\ref{JLAandPan}.
For $1\sigma$ CL, comparing to the case of using JLA data,
using Pantheon data will decrease the the error bars of EoS $w$ by 45\%, 43\% and 56\%, for MS, FS and IFS, respectively.
For $2\sigma$ CL, compared with the case of using JLA data,
using Pantheon data will decrease the the error bars of EoS $w$ by 38\%, 47\% and 53\%, for MS, FS and IFS, respectively.
In addition, using Pantheon data can also increase the values of FoM by 373\%, 127\% and 153\%, for MS, FS and IFS, respectively.
These results show that compared with the JLA data, the Pantheon data can provide the much tighter DE constraints.

\subsection{The difference among the cosmological consequences of MS, FS and IFS}
In this part, we compare the difference among the cosmological consequences of MS, FS and IFS.

\begin{figure}[htbp]
      \centering

      \includegraphics[width=7cm,height=5cm,trim=120 270 120 270,clip]{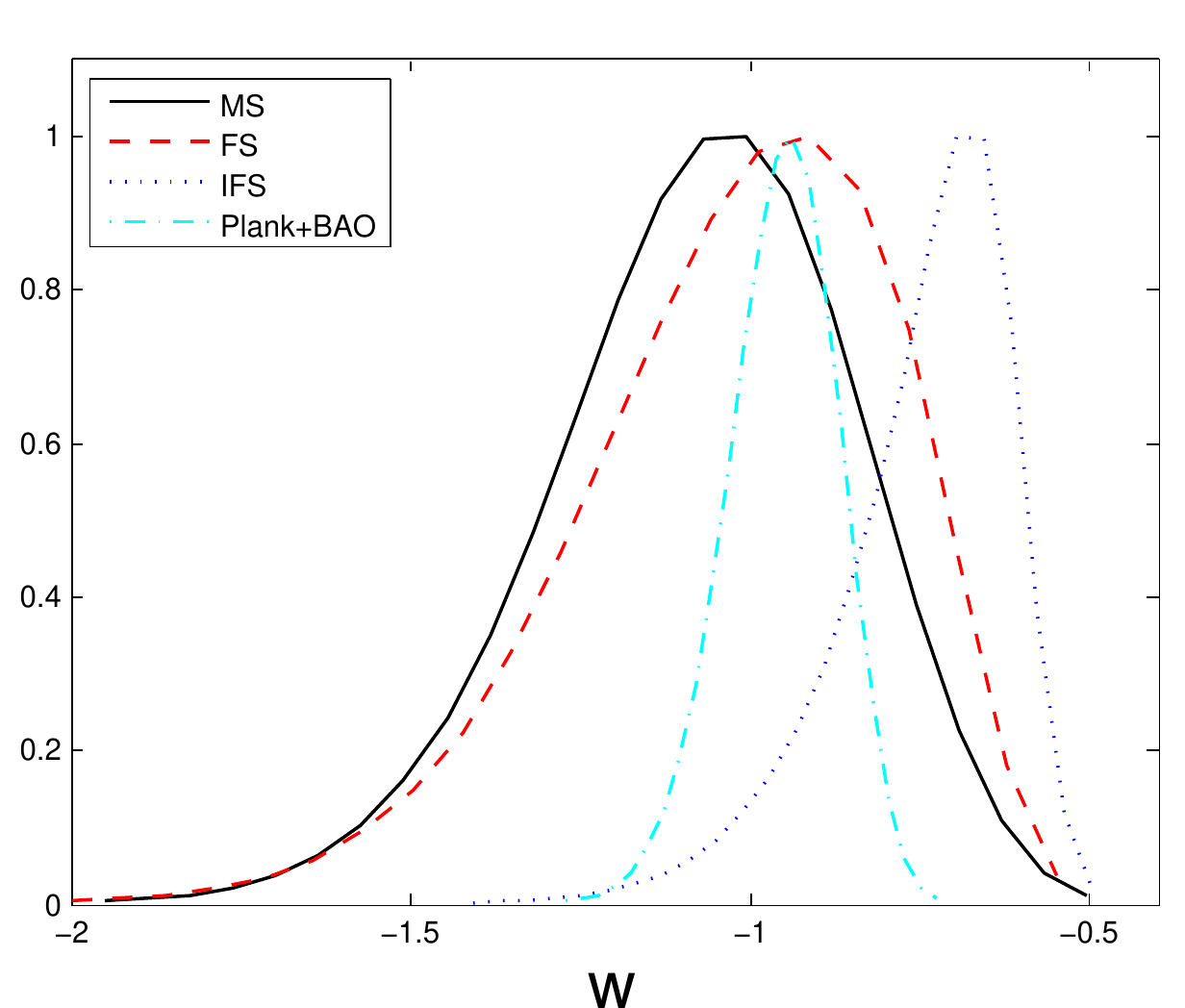}

      \caption{The 1D marginalized probability distributions of $w$ produced by the Pantheon data-set. The solid black, dashed red and dotted blue lines denote the results given by MS, FS and IFS, respectively.
       As a comparison, we also constrain the $wCDM$ model by adopting the combined CMB+BAO data, which is represented by the dashdotted cyan line.}
      \label{fig:wcompare}
    \end{figure}

In Fig.\ref{fig:wcompare}, we present the 1D marginalized probability distributions of $w$.
The solid black, dashed red and dotted blue lines denote the results given by MS, FS and IFS, respectively.
As a comparison, we also constrain the $wCDM$ model by adopting the combined CMB+BAO data, which is represented by the dashdotted cyan line.
From this figure, one can see that the IFS can give the tightest constraint among three SN Ia statistics techniques.
In addition, comparing with other SN Ia statistics techniques, FS yields a more similar marginalized probability distribution of $w$ to that given by CMB+BAO data-set.

\begin{figure}[htbp]
      \centering

      \includegraphics[width=7cm,height=5cm,trim=20 0 20 20,clip]{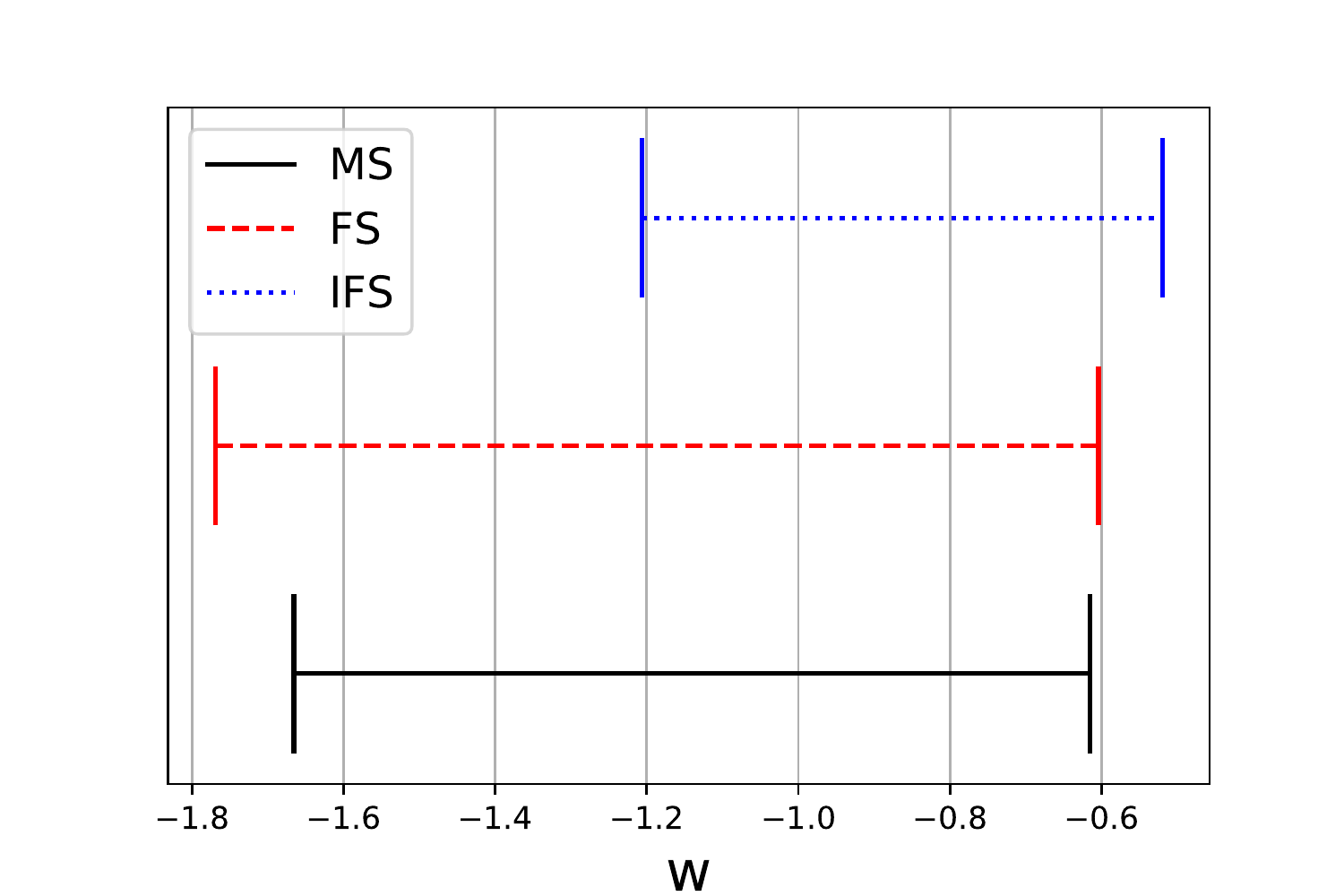}
      \caption{The $2\sigma$ error bars of EoS $w$ given by three SN Ia statistics techniques. The solid black, dashed red and dotted blue lines denote the results given by MS, FS and IFS, respectively. It's clear that the IFS yields the tightest constraint on $w$.}
      \label{fig:error}
      \end{figure}

In Fig.\ref{fig:error}, we plot the $2\sigma$ error bars of $w$ for three SN Ia statistics techniques.
The solid black, dashed red and dotted blue lines represent the results of MS, FS and IFS, respectively.
One can find that, using IFS will yield the tightest constraint on $w$.
Moreover, using FS will give a smallest lower limit of EoS $w$, which is less than $-1$.
As will be discussed in the next subsection, this will lead to a ``cosmic doomsday''.

  \begin{table*}
    \caption{\label{JLAandPan} The FoM and CL of $w$ for JLA and Pantheon data-set  }\centering
    \begin{tabular*}{\textwidth}{@{}l*{15}{@{\extracolsep{0pt plus12pt}}l}}
    \hline\hline

    Parameters&MS    &FS   &IFS \\
    \hline
    The $1\sigma$ CL of $w_{JLA}$ &( -1.485 , -0.311 )&( -1.965 , -0.513 ) &( -1.426 , -0.457 )\\
    The $1\sigma$ CL of $w_{Pantheon}$&( -1.407 , -0.761 )&( -1.460 , -0.634 )&( -0.965 , -0.547 )      \\
    The decrease of $1\sigma$ error bar& 45\%&43\%& 56\%    \\
    \hline
    The $2\sigma$ CL of $w_{JLA}$ &( -1.903 , -0.218 )&( -2.667 , -0.459 ) &( -1.878 , -0.426 )\\
    The $2\sigma$ CL of $w_{Pantheon}$&( -1.666 , -0.615 )&( -1.769 , -0.604 )&( -1.206 , -0.520 )      \\
    The decrease of $2\sigma$ error bar& 38\%&47\%& 53\%    \\
    \hline
    $FoM$ of JLA&34.2&61.2 &109.8\\
    $FoM$ of Pantheon&161.8&139.2 &278.3\\
    The increase of  $FoM$&373\%&127\%& 153\%\\
    \hline
    \end{tabular*}
    \end{table*}
\subsection{The fate of the Universe}
\begin{table*}
    \caption{\label{riptime} The time remaining before``big rip'' for three SN Ia statistics techniques }\centering
    \begin{tabular*}{\textwidth}{@{}l*{15}{@{\extracolsep{0pt plus12pt}}l}}
    \hline\hline

    $t_{rip}-t_0$&MS    &FS   &IFS \\
    \hline
    the best-fit prediction& 255.5Gyr&N/A& N/A\\
    the $2\sigma$ low limit prediction& 19.3Gyr&19.0Gyr& 116.6Gyr    \\
    \hline
    \end{tabular*}
    \end{table*}

In this subsection, based on the cosmology-fit results obtained above, we discuss the fate of the Universe.

As shown in Ref.\cite{PhysRevLett.91.071301}, a phantom DE with EoS $w<-1$ will cause a ``cosmic doomsday''.
The reason is that the energy density of phantom DE will increase along with time $t$.
This means that the repulsive force of phantom DE will also increase along with $t$.
Therefore, sooner or later, the repulsive force of phantom DE will rip apart all the structures in the Universe.
This is so called ``big rip''.

If the Universe is dominated by the phantom DE,
the time remaining before the Universe ends can be calculated as \cite{PhysRevLett.91.071301}
\be\label{trip}
t_{rip}-t_0\simeq\frac{2}{3}\frac{1}{|1+w|H_0\sqrt{1-\Omega_m}},
\ee
where $t_{rip}$ is the time of  ``big rip'', while $t_0$ is current time.

In Table.\ref{riptime}, we list the results of time interval $t_{rip}-t_{0}$ corresponding to the best-fit points and $2\sigma$ lower limit
of MS, FS and IFS fitting results, respectively.
For the worst case (i.e. $2\sigma$ lower limit), the results of $t_{rip}-t_{0}$ are 19.3, 19.0, 116.6 Gyr for MS, FS, and IFS, respectively.
In addition, MS gives a best-fit point $(\Omega_m,w)=(0.314,-1.05)$, which yields a $t_{rip}-t_{0}=255.5$ Gyr.
On the other hand, FS and IFS give the best-fit points $(\Omega_m,w)=(0.273,-0.96)$ and $(0.100,-0.61)$, respectively,
which lead to a Universe without ``cosmic doomsday''.
This means that, MS more favors a Universe that will encounter a  ``big rip''.

Now, Let us discuss the topic of ``big rip'' with more details.
For a gravitationally bound system with mass M and radius R, the period of a circular orbit around this system at
radius R is $P = 2\pi(R^3/GM)^{1/2}$, where G is the Newton¡¯s constant.
Along with the increase of the repulsive force of phantom DE, sooner or later, this system will become unstable.
This means that, in future, at a moment $t_{tear}$, this  gravitationally bound system will be destroyed by phantom DE.
As pointed out in Refs. \citep{PhysRevLett.91.071301,Li:2012via},
$t_{rip}$ and $t_{tear}$ satisfy the relation
\be
t_{rip}-t_{tear}\simeq P\frac{\sqrt{2|1+3w|}}{6\pi|1+w|}.
\ee

\begin{figure}[htbp]
      \centering

      \includegraphics[width=7cm,height=5cm,trim=28 10 40 50,clip]{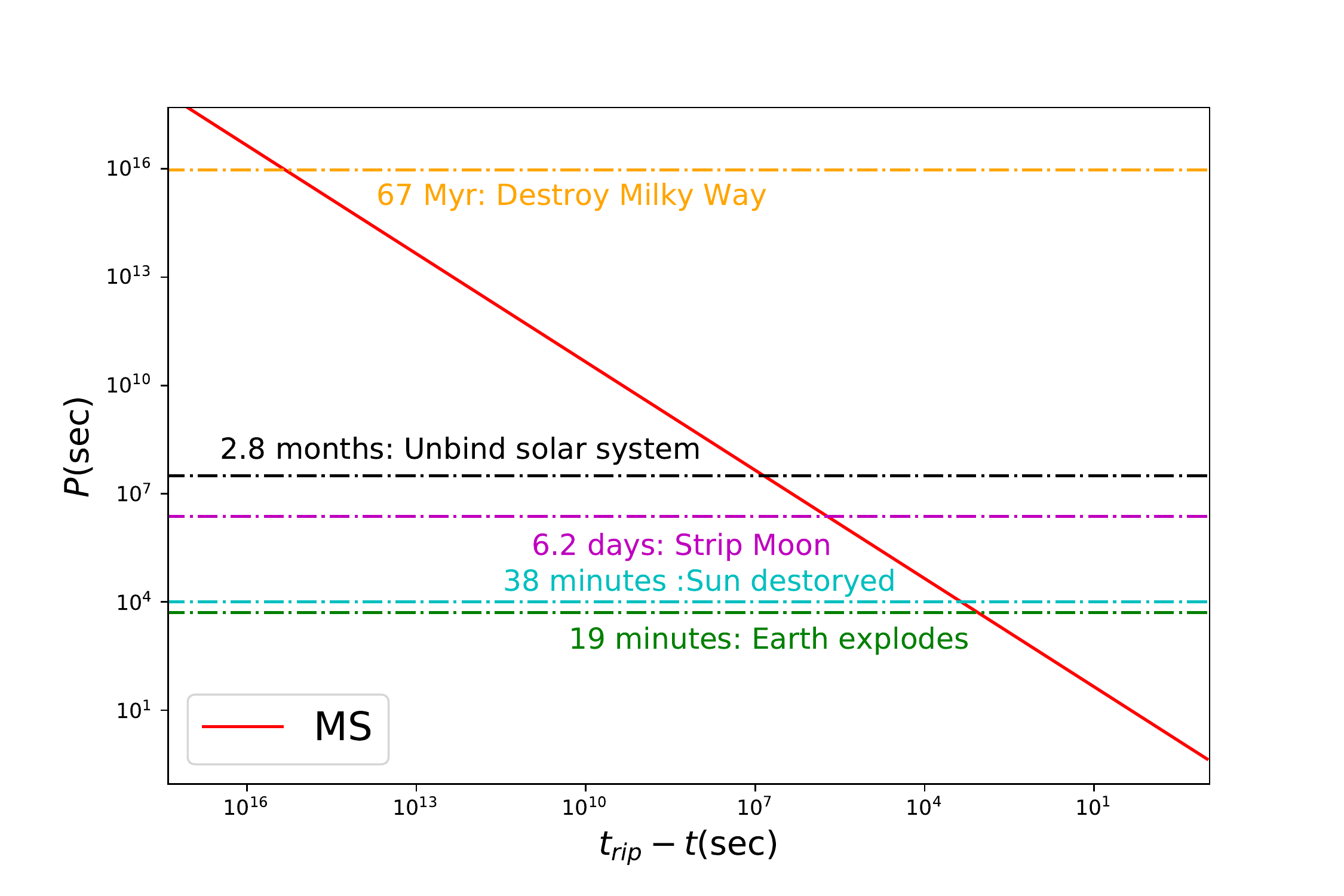}
      \caption{The relation between P and the time of system stripped before ``big rip''. These results are calculated by adopting $2\sigma$ lower limit results of MS, where the Pantheon data-set are used in the cosmology-fits.
      The dashdotted horizontal lines denote different $P$ for some characteristic structures such as the Milky Way,
       the solar system, the Earth-moon system, the Sun, and the Earth.
       }
      \label{fig:P_t}
      \end{figure}

In Fig.\ref{fig:P_t}, by adopting the $2\sigma$ lower limit results of MS, we plot the relation
between P and $t_{rip}-t_{tear}$ for some characteristic structures.
It should be mentioned that the Pantheon data-set are used in the cosmology-fits.
From this figure, one can see that
the Milky Way will be destroyed 67 Myr before the big rip;
2.8 months before the doomsday, the Earth will be ripped from the Sun;
6.2 days before the doomsday, the moon will be ripped from the Earth;
the Sun will be destroyed 38 minutes before the end of time;
and 19 minutes before the end, the Earth will explode.
At the moment of big rip $t_{rip}$, everything, including microscopic object such as atom,
will be torn apart by the repulsive force of phantom DE.

\section{Conclusions and Discussions}\label{conclu}
In this work, we investigate the cosmological consequences of the latest SN Ia data-set, Pantheon, by using the $wCDM$ model.
As a comparison, we also consider the JLA data-set.
Moreover, three kinds of SN Ia statistics techniques, including MS, FS and IFS, are taken into account.
In addition, we mainly focus on the cosmological consequences given by using SN data alone.

First, based on the $wCDM$ model, we scan the parameter space of $(z_{cut},\Delta z)$ to determine the best recipe for IFS.
Then, we compare the difference between the fitting results given by the Pantheon and JLA data.
Next, by using SN Ia samples alone, we compare the difference among the cosmological consequences given by MS, FS and IFS.
Finally, based on the cosmology-fit results obtained above, we discuss the fate of the Universe.

Our main results are showed as follows:
\begin{itemize}
 \item
  For IFS, We find the best FA recipe $(z_{cut},\Delta z)=0.1,0.08)$, which can give the largest $FoM=278.34$ (see Fig.\ref{fig:scan}).
  Comparing to the case of adopting the combined SN+CMB+BAO data-set,
  using SN data alone will yield a smaller value of $z_{cut}$ (see Fig.\ref{fig:cut}).
 \item
  Comparing to the JLA data-set, the Pantheon data-set can give tighter DE constraints (see Fig.\ref{fig:three}).
  Specifically, the Pantheon data can decrease the $2\sigma$ error bars of $w$ by $38\%$ $47\%$ and $53\%$, for MS, FS and IFS, respectively.
  In addition, using Pantheon data can also increase the values of FoM by 373\%, 127\% and 153\%, for MS, FS and IFS, respectively (see Table.\ref{JLAandPan}).
 \item
  FS gives closer results to other observations, such as BAO and CMB (see Fig.\ref{fig:wcompare}).
  In addition, among three SN Ia statistics techniques, IFS yields the tightest constraint on $w$ (see Fig.\ref{fig:error}).
 \item
  For the case of adopting best-fit results, MS yields a $t_{rip}-t_{0}=255.5$ Gyr, while FS and IFS favor a Universe that will expend eternally.
  This means that MS more favors a Universe that will end in a ``big rip''(see Table.\ref{riptime}).
  In addition, we also show the specific moments those various characteristic structures are ripped by the phantom DE (see Fig.\ref{fig:P_t}).
\end{itemize}

In this paper, we only discuss a specific DE model, i.e. the $wCDM$, which has a constant EoS $w$.
It is interesting to consider the cases of adopting various dynamics DE models, such as
quintessence \citep{Zlatev:1998tr}, Chaplygin gas \citep{Kamenshchik:2001cp}, holographic DE \citep{Li:2004rb}, agegraphic DE \citep{Wei:2007ty}
, Yang-Mills condensate \citep{Wang:2008fx},  Chevalliear-Polarski-Linder parametrization \citep{Chevallier:2000qy,Linder:2002et} and  binned parametrization \citep{Huang:2009rf,Wang:2010vj,Li:2011wb}.

In addition, in this study we only compare the cosmological consequences of the SN observation with the CMB and BAO observations.
It's also interesting to  compare the results of the SN observation with some other cosmological observations,
such as weak gravitational lensing \citep{Blandford:1991edc}, abundance of galaxy clusters \citep{Allen:2011zs}, Alcock-Paczynski effect \citep{Alcock:1979mp}, direct $H_0$ measurement \citep{Hu:2004kn}, and cosmic age test \citep{Wang:2008te,Wang:2010su}. These will be done in future works.

\begin{acknowledgments}
 SW is supported by the National Natural Science Foundation of China under Grant No. 11405024 and the Fundamental Research Funds for the Central Universities under Grant No. 16lgpy50.
\end{acknowledgments}



\bibliography{ref}
\end{document}